\documentclass[prl,twocolumn, notitlepage,superscriptaddress, nofootinbib]{revtex4-1}

\usepackage{fancyhdr}
\fancyheadoffset[L,R]{\headrulewidth}
\renewcommand{\headrulewidth}{0.6pt}

\usepackage{cancel}
\usepackage[normalem]{ulem}

\usepackage{cancel}

\usepackage{graphicx}
\usepackage{tikz,pgf}
\usepackage{color}
\usepackage{array}
\usepackage{xcolor}
\usepackage[dvips]{epsfig}
\usepackage{epsfig}
\usepackage{enumerate}
\usepackage[latin1]{inputenc}
\usepackage[T1]{fontenc}
\usepackage[hang]{subfigure}
\usepackage{bbm, dsfont}
\usepackage{amsfonts,amsmath,amssymb,stmaryrd}
\usepackage{ulem}

\newcommand{\bd}{\hat b^\dagger}

\usepackage{bbold}

\usepackage{simplewick}

\newcommand{\bra}[1]{\langle #1 | \,}
\newcommand{\ket}[1]{\, | #1 \rangle}
\newcommand{\braket}[2]{\langle #1 | #2 \rangle}

\newcommand{\ga}{\ga}

\newcommand{\bl}{\begin{linenomath*}}
\newcommand{\el}{\end{linenomath*}}

\newcommand{\bea}{\begin{eqnarray}}
\newcommand{\eea}{\end{eqnarray}}

\renewcommand{\a}{\hat a}
\newcommand{\ad}{\hat a^\dagger}

\newcommand{\be}{\hat b}
\newcommand{\bed}{\hat b^\dagger}
\renewcommand{\ga}{\hat\gamma}

\newcommand{\veck}{\mathbf k}

\definecolor{dgreen}{rgb}{0.0, 0.5, 0.0}

\usepackage{hyperref}

\begin{document}

\title{Rotation of quantum impurities in the presence of a many-body environment}

\author{Richard Schmidt} 
\email{richard.schmidt@cfa.harvard.edu}
\affiliation{ITAMP, Harvard-Smithsonian Center for Astrophysics, 60 Garden Street, Cambridge, MA 02138, USA}%
\affiliation{Physics Department, Harvard University, 17 Oxford Street, Cambridge, MA 02138, USA} %

\author{Mikhail Lemeshko} 
\email{mikhail.lemeshko@ist.ac.at}
\affiliation{IST Austria (Institute of Science and Technology Austria), Am Campus 1, 3400 Klosterneuburg, Austria}

\begin{abstract}
We develop a microscopic theory describing a quantum impurity whose rotational degree of freedom is coupled to a  many-particle bath. We approach the problem by introducing the concept of an ``angulon'' -- a quantum rotor dressed by a quantum field -- and reveal its quasiparticle properties using a combination of variational and diagrammatic techniques. Our theory predicts renormalisation of the impurity rotational structure, such as observed in experiments with molecules in superfluid helium droplets, in terms of a rotational Lamb shift induced by the many-particle environment. Furthermore, we discover a rich many-body-induced fine structure, emerging in rotational spectra due to a redistribution of angular momentum within  the quantum many-body system.

\end{abstract}

\maketitle

The concepts of rotation and angular momentum are ubiquitous across quantum physics, whether one deals with the lifetimes of unstable nuclei~\cite{WaleckaNuclearBook},  accuracy of atomic clocks~\cite{DereviankoRMP11}, or electronic structure of defect centers in solids~\cite{MazeNJP11}. Pioneered by the seminal works of Wigner~\cite{WignerGroupTheory} and Racah~\cite{RacahComplSpec1, *RacahComplSpec2, *RacahComplSpec3}, the quantum theory of angular momentum evolved into a powerful machinery, commonly used to classify the states of isolated quantum systems and perturbations to their structure due to electromagnetic or crystalline fields~\cite{VarshalovichAngMom, ZareAngMom}.  
In ``realistic'' experiments, however, quantum systems are almost inevitably coupled to a many-particle environment and associated fields of excitations, which is capable of profoundly altering the physics of the system~\cite{AbrikosovBook, WeissBook, Breuer2002}.

Although studying the effects of a fluctuating bath on the dynamics of quantum impurities represents a vast research field of its own~\cite{WeissBook, Breuer2002}, an understanding of the many-body effects in the context of rotational degrees of freedom is still at its embryonic stage.   As opposed to translational motion, whose coupling to a quantum field constitutes the well-studied ``polaron problem''~\cite{LandauPolaron,LandauPekarJETP48,AppelPolarons, Devreese13}, quantum rotations in three-dimensional space are described by a non-Abelian $SO(3)$ algebra and possess a discrete spectrum of eigenenergies. This results in a tremendous complexity of the angular momentum properties even for a few interacting particles~\cite{VarshalovichAngMom}.
 On the other hand, unlike spin degrees of freedom~\cite{LeggettRMP87}, rotation is explicitly associated with an intrinsic motion of the system, whose coupling to a bath needs to be properly accounted for by a microscopic theory. As a consequence, the description of rotating particles  immersed into a many-body system cannot be reduced to any of the previously known impurity problems of condensed matter physics.

Here we uncover rich physics associated with quantum rotation coupled to a many-particle environment. Using a combination of variational and  diagrammatic techniques, we demonstrate that the problem can be described within the quasiparticle picture of an ``angulon'' -- a quantum rotor dressed by a quantum many-body field. The angulon is a collective object, characterized by the total angular momentum of the system, of which it is an eigenstate. At the same time, due to the impurity-bath interactions, the angular momentum is shared within the many-particle system, which results in a peculiar behavior of the angular momentum eigenstates. In particular, we discover the rotational Lamb shift as well as a rich many-body-induced fine structure emerging due to the transfer of angular momentum between the impurity and the many-particle bath.  The latter effect has no direct analogue in isolated atoms and molecules. We identify the regimes where the predicted effects are accessible by experiments on cold molecules in Bose-Einstein Condensates (BEC's)  and  inside superfluid helium nanodroplets.

As a first step, we derive the effective Hamiltonian of the angulon problem, starting from a microscopic description.  For the sake of clarity, we exemplify the derivation by considering a linear-rotor molecule immersed into a weakly-interacting BEC. The resulting Hamiltonian, Eq.~(\ref{Hamil}) is, however,  applicable to a broad range of systems consisting of a quantum rotor coupled to a many-body bath of harmonic oscillators, such as phonons in a liquid~\cite{ToenniesAngChem04} or solid \cite{Mahan90}, as well as microwave photons or electromagnetic field noise~\cite{GardinerZollerBook}.

The Hamiltonian for a molecular impurity in a homogeneous BEC is given by $H =  H_\text{bos} + H_\text{imp} + H_\text{imp-bos}$, where $H_\text{bos}= \sum_\mathbf{k}  \epsilon_\veck \ad_\mathbf{k} \a_\mathbf{k} +g_\text{bb} \sum_\mathbf{k, k', q} \ad_\mathbf{k'-q} \ad_\mathbf{k+q} \a_\mathbf{k'}  \a_\mathbf{k}$ describes a gas of bosons with a dispersion relation $\epsilon_\veck$, whose contact interactions are given by the strength $g_\text{bb}>0$ \cite{Pitaevskii2003}. In units where $\hbar \equiv 1$, the kinetic energy of a linear rotor impurity is given by $H_\text{imp} = B \mathbf{\hat{J}^2}$, where $B$ is the rotational constant and $\mathbf{\hat{J}}$ is the angular momentum operator.  Accordingly, the non-interacting impurity eigenstates, $\vert j, m \rangle$, are characterized by the angular momentum, $j$, and its projection, $m$, onto the laboratory-frame $z$-axis, and form $(2j+1)$-fold degenerate multiplets with energies $E_j = B j(j+1)$~\cite{LevebvreBrionField2, LemKreDoyKais13}. We consider a molecule whose linear motion is frozen, which is a good approximation for the experimentally relevant cases~\cite{ToenniesAngChem04}.

\begin{figure}[t]
  \centering
 \includegraphics[width=\linewidth]{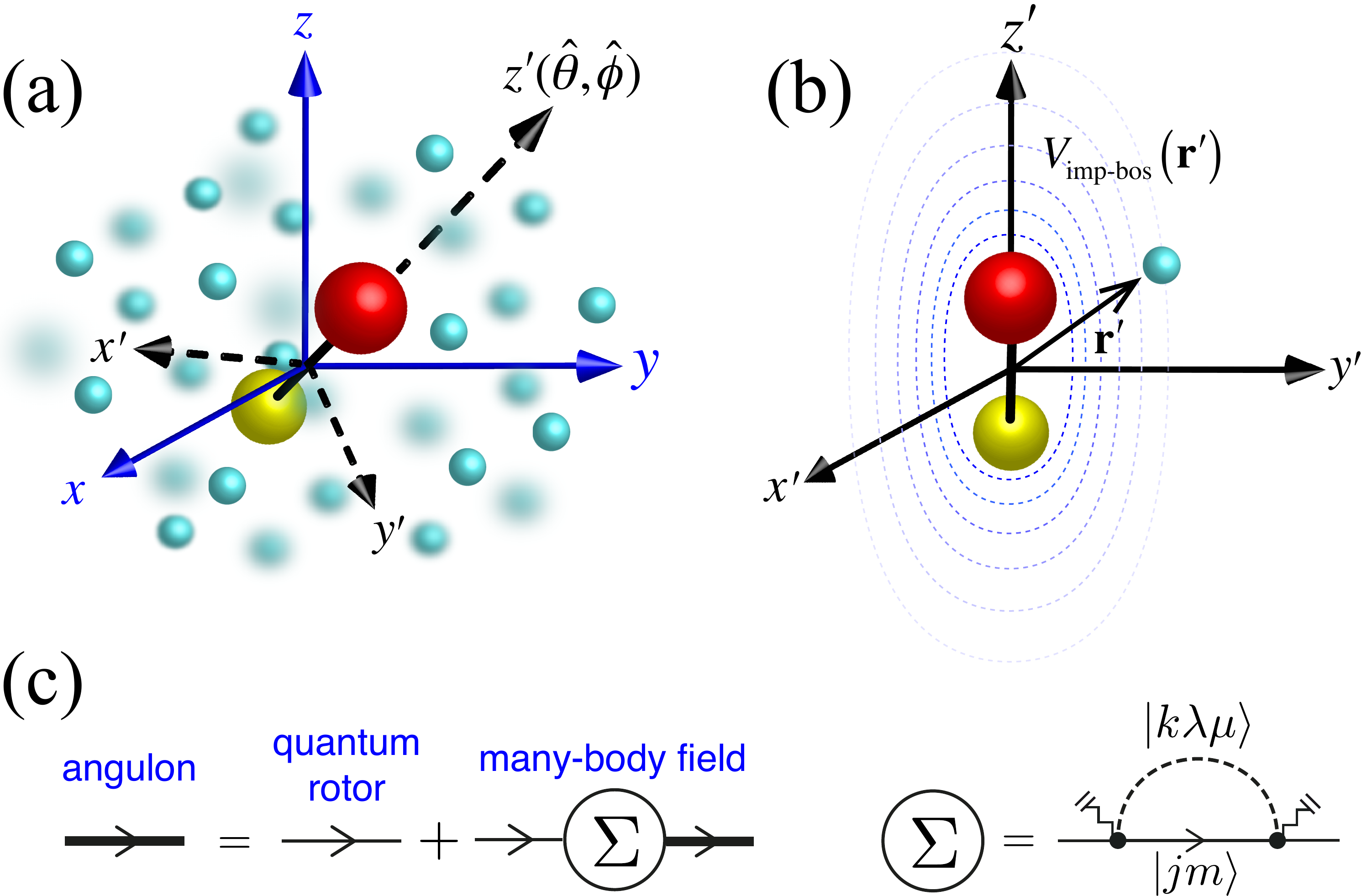}
  \caption{\label{rotor} (a) The interaction of a quantum rotor with a quantum many-body system explicitly depends on the rotor angular coordinates, $(\hat{\theta}, \hat{\phi})$, in the laboratory frame. (b) The anisotropic rotor-boson interaction is defined in the rotor coordinate frame, $\mathbf{r'} = (r',\Theta', \Phi')$. (c) Representation of Eq.~\eqref{GFct} in terms of Feynman diagrams.}
 \end{figure}

Anisotropic molecular geometry gives rise to anisotropic impurity-boson interactions, 
\begin{equation}
\label{Himpbos}
H_\text{imp-bos} =\sum_\mathbf{k, q}  V_\text{imp-bos} (\mathbf{q}, \hat{\theta}, \hat{\phi})  \ad_\mathbf{k-q} \a_\mathbf{k},
\end{equation}
which are explicitly dependent on the molecular orientation in the laboratory frame, as defined by the angles $(\hat\theta,\hat\phi)$, see~Fig.~\ref{rotor}(a).  Here the two-body interaction, $V_\text{imp-bos} (\mathbf{q}, \hat{\theta}, \hat{\phi}) = \mathcal{F} [R(\hat{\theta}, \hat{\phi}) V_\text{imp-bos} (\mathbf{r'}) ]$, is obtained by applying the rotation operator, $R(\hat{\theta}, \hat{\phi})$, to the molecular-frame potential, $V_\text{imp-bos} (\mathbf{r'})$, and $\mathcal F$ denotes the subsequent Fourier transform. For a linear rotor,  the anisotropic  potential can be expanded in spherical harmonics, $V_\text{imp-bos} (\mathbf{r'}) = \sum_\lambda u_\lambda f_\lambda (r') Y_{\lambda 0} (\Theta', \Phi')$, where  $(r',\Theta', \Phi')$ are the boson coordinates in the molecular frame and $u_\lambda$ and $f_\lambda(r')$ represent the strength and shape of the potential in the respective angular momentum channel $\lambda$, cf.~Fig.~\ref{rotor}(b) \cite{sup}.

After expanding the Hamiltonian in fluctuations around a homogeneous BEC of condensate density $n$~\cite{Wang2005,Wang2006,TemperePRB09}, its bosonic part can be diagonalised using the Bogoliubov transformation~\cite{Pitaevskii2003}. The resulting theory is expressed in terms of bosonic quasiparticles (`phonons') with dispersion relation $\omega_k$. The corresponding creation and annihilation operators are  conveniently expressed  in the spherical basis, $\bd_{k\lambda \mu} = k (2\pi)^{-3/2} \int d\Omega_k~\bd_\mathbf{k} ~i^\lambda~ Y^*_{\lambda \mu} (\Omega_k)$. Here $k=|\mathbf{k}|$, and $\lambda$ and $\mu$ define, respectively, the phonon angular momentum and its projection onto the $z$-axis. Up to a constant mean-field shift, the resulting effective Hamiltonian reads
\begin{multline}
\label{Hamil}
H= B \mathbf{\hat{J}^2} + \sum_{k \lambda \mu}  \omega_k \bed_{k\lambda \mu} \be_{k\lambda \mu} \\+   \sum_{k \lambda \mu} U_\lambda(k)  \left[ Y^\ast_{\lambda \mu} (\hat \theta,\hat \phi) \bed_{k \lambda \mu}+ Y_{\lambda \mu} (\hat \theta,\hat \phi) \be_{k \lambda \mu} \right],
\end{multline}
with $\sum_k\equiv\int dk$ and the angular momentum dependent interaction given by $U_\lambda(k) = u_\lambda \left[\frac{8 n k^2\epsilon_k}{\omega_k(2\lambda+1)}\right]^{1/2} \int dr r^2 f_\lambda(r) j_\lambda (kr)$ with $j_\lambda(kr)$ the spherical Bessel function \cite{sup}. 

It is important to note that, although we derived Eq.~\eqref{Hamil} in the context of a particular physical realisation, an effective Hamiltonian of the same structure can be obtained for any quantum system whose angular momentum is coupled to a bath   of quantum oscillators. Having this in mind,  we will hereafter regard  Eq.~(\ref{Hamil}) from a  general perspective  and study its generic properties both within and beyond the cold-atom regime.
While the first two terms of Eq.~\eqref{Hamil} correspond to the bare kinematics of the impurity and the bath, the last term accounts for the absorption and emission of field quanta by the quantum rotor. As opposed to the spin-boson model~\cite{LeggettRMP87}, this term explicitly depends on the molecular angle operators, $(\hat \theta,\hat \phi)$, which is essential for the microscopic description of physical rotation and the emergence of angulon physics. On the other hand, the spherical harmonics appearing in Eq.~\eqref{Hamil} incorporate the non-Abelian algebra of angular momentum, which is not present in the Fr\"ohlich Hamiltonian for a translationally moving impurity~\cite{Devreese13}.

While exact solutions for Eq.~\eqref{Hamil} exist  in the limit of a non-rotating molecule, this is no longer the case for finite  $B$, where the rotor becomes a dynamical degree of freedom.  In order to get insight into the angulon properties, including the non-perturbative regime, we introduce a variational ansatz for the many-body quantum state which is based on an expansion in  bath excitations,
\begin{equation}
\label{VarFunc}
	\vert \psi \rangle = Z_{LM}^{1/2} \ket{0} \ket{L M}+ \sum_{\substack{k \lambda \mu \\ j m}} \beta^{LM}_{k \lambda j} C_{jm, \lambda \mu}^{L M} \bed_{k \lambda \mu} \ket{0} \ket{jm},
\end{equation}
where $\ket{0}$ represents the vacuum of bath excitations  and the angulon quasiparticle weight is given by the normalisation condition, $Z_{LM}\equiv 1- \sum_{k \lambda j} |\beta^{LM}_{k \lambda j}|^2 $. 
The many-body state (\ref{VarFunc}) is an eigenstate of the total angular momentum, $\mathbf{\hat{L}^2} \vert \psi \rangle = L(L+1)\vert \psi \rangle $, and its projection on the laboratory $z$-axis, $\hat{L}_z \vert \psi \rangle = M\vert \psi \rangle$, which is incorporated by the Clebsch-Gordan coefficient $C_{jm, \lambda \mu}^{L M}$~\cite{VarshalovichAngMom, ZareAngMom}. In the absence of external fields, the   quantum number $M$ is irrelevant and will be omitted hereafter.

\begin{figure*}[t]
  \centering
 \includegraphics[width=\linewidth]{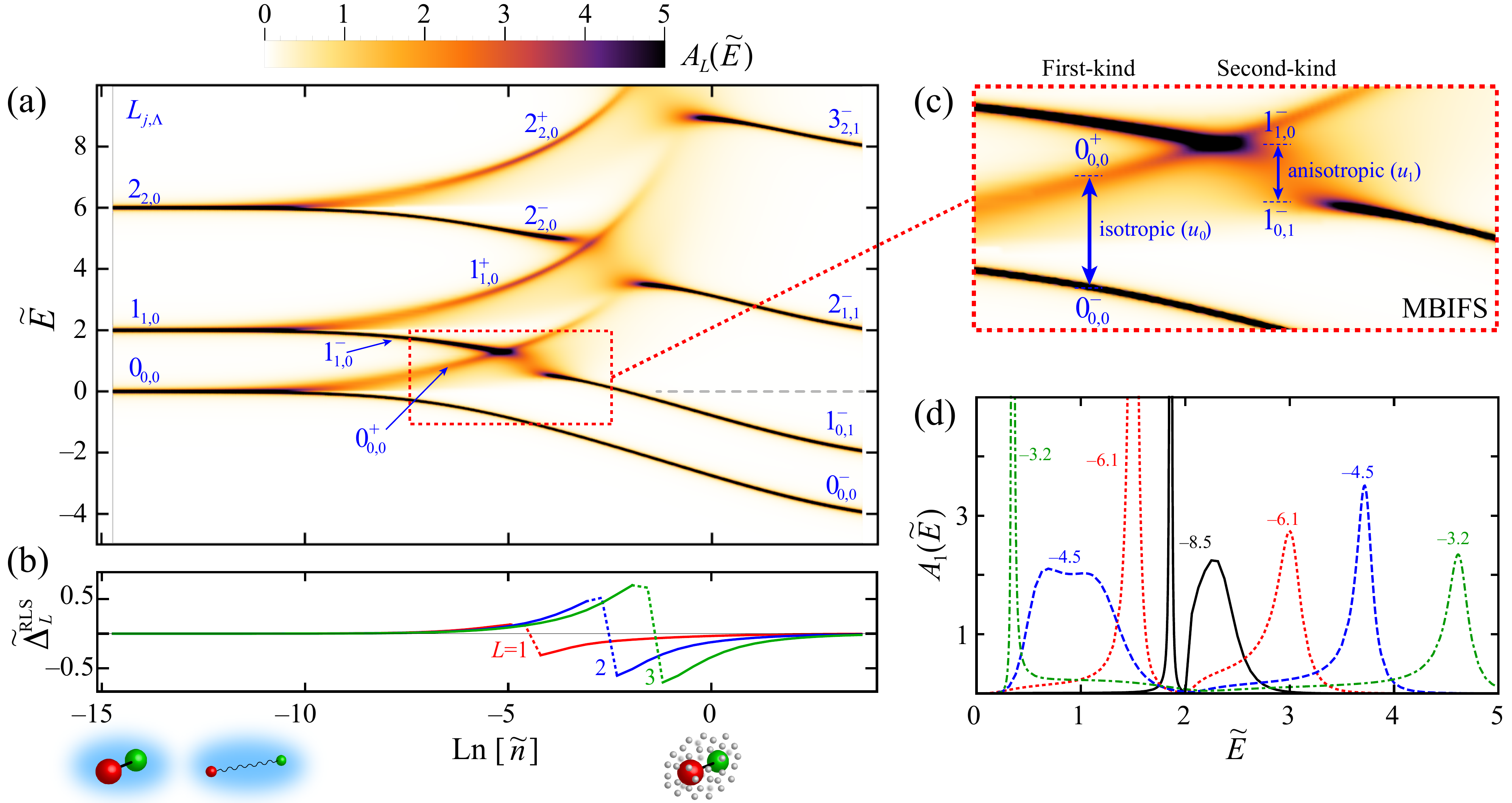}\vspace{-3mm}
  \caption{\label{Stark} (a) The angulon spectral function, $A_L (\tilde{E})$, as a function of the dimensionless density, $\tilde{n}=n (m B)^{-3/2}$, and energy $\tilde E=E/B$. The low- and high-density regimes can be realized with cold molecules trapped inside a BEC and superfluid helium droplets, respectively. The ground state, $0_{00}^-$, is stable; so is the state $1_{01}^-$ after it crosses the phonon threshold at zero energy, marked by the horizontal dashed line. Within the plotted range of $\tilde{n}$, the rest of the excited states have finite lifetimes. (b)~Differential rotational Lamb shift for the lowest nonzero-$L$ states. (c)~Zoom-in illustrating the Many-Body-Induced Fine Structure (MBIFS) of the first kind, $L_{L,0} \to \{L_{L, 0}^-, L_{L, 0}^+\}$, and of the second kind, $L_{L,0}^- \to L_{L-1,1}^-$. (d) Spectroscopic signatures of the MBIFS for the $L=1$ state. The numbers indicate the corresponding values of $\text{ln}[\tilde{n}]$.  Sharp features in panels (a), (c), and (d) were artificially broadened for better visibility. \vspace{-0.3cm}}
 \end{figure*}
Minimization of  the energy, $E = \bra{\psi}H\ket{\psi}/ \langle \psi \vert \psi \rangle$, subject to the constraint $\braket{\psi}{\psi}=1$, yields the variational energies, $E_L$, as solutions of the self-consistent equation, $\left[G^\text{ang}_L(E_L)\right]^{-1}=0$, where
\begin{equation}
\label{GFct}
\left[G^\text{ang}_L(E)\right]^{-1}= [G^0_L(E)]^{-1} - \Sigma_L(E) ,
\end{equation}
with $[G^0_L(E)]^{-1} =-E+BL (L+1)$ and
\begin{equation}
\label{Selfenergy}
\Sigma_L (E) =  \sum_{k \lambda j} \frac{2\lambda+1}{4\pi} \frac{U_\lambda (k)^2 \left[C_{L0, \lambda0}^{j0} \right]^2 }{B j(j+1) - E+ \omega_k}.
\end{equation}
Eq.~\eqref{GFct}  is equivalent to the Dyson equation $G^\text{ang}= G^0 +G^0 \Sigma G^\text{ang}$ for the interacting Green's function, $G^\text{ang}_L(E)$, with the self-energy, $\Sigma_L(E)$, calculated from the diagrammatic expansion shown in Fig.~\ref{rotor}(c). As a consequence, our variational technique is equivalent to a diagrammatic approach to the problem. As such, it allows to access not only the ground-state properties but also the entire excitation spectrum of the system by virtue of the retarded Green's function, $G_L^\text{ret}(E)=G_L^\text{ang}(E+i0^+)$. The  corresponding self-energy is evaluated in closed form:
\begin{eqnarray}
 \label{ImSigma}
  \text{Im} \Sigma^\text{ret}_L(E)&=&\sum_{\lambda j k_0}\theta(E-Bj(j+1))\left[C_{L0, \lambda0}^{j0} \right]^2 \nonumber\\ &&\times\,\, \frac{2\lambda+1}{4} U_\lambda (k_0)^2 \vert(\partial \omega_k/\partial k)_{k=k_0}\vert^{-1},
 \end{eqnarray}
with $\text{Re}\Sigma^\text{ret}_L(E)$ given by the Kramers-Kronig relations~\cite{Mahan90}. Here $k_0$ gives the roots of $E-\omega_k+Bj(j+1) =0$. The theta-function represents the onset of  phonon bands, shown in Fig.~\ref{Stark}(d) (see below), which have been observed in experiments with helium nanodroplets~\cite{Hartmann2002}.   The analytic results allow for the efficient calculation of  the spectral function,  $\mathcal A_L (E)= \text{Im}[G^\text{ret}_L(E)]$, which encompasses both the  excitation spectrum as well as  the quasiparticle properties of the angulon, and therefore is the central object to be studied below.

Let us analyze the generic properties of Eq.~\eqref{Hamil}, by considering a molecule immersed in a superfluid with  dispersion relation $\omega_k=\sqrt{\epsilon_k(\epsilon_k+2 g_{\text{bb}} n)}$, where $g_\text{bb}=4\pi a_{\text{bb}}/m$ \cite{Pitaevskii2003}. Real atom-molecule potentials depend on the particular system and comprise high-order terms in their spherical harmonics expansion~\cite{StoneBook13}.  In Eq.~\eqref{Hamil},  each   term of the expansion leads to an effective phonon-mediated coupling between the bare rotational states. In order to keep the analysis of the Hamiltonian most transparent, we focus on the leading terms, $\lambda=0,1$, which suffices to reveal the intricate interplay between the anisotropic and isotropic interaction channels  responsible for the emergent properties of the angulon.

In Fig.~\ref{Stark}(a) we show the resulting angulon spectral function, $A_L(\tilde{E}) \equiv \mathcal{A}_L(\tilde{E}) B$, as a function of the dimensionless superfluid density, $\tilde{n}=n (m B)^{-3/2}$  and energy $\tilde{E} = E/B$. We choose Gaussian form factors, $f_\lambda(r) = (2\pi)^{-3/2} e^{-r^2/(2r_\lambda^2)}$, and interaction parameters $u_0 = 1.75 \, u_1 = 218 B$,  $r_0 = r_1 = 1.5\, (m B)^{-1/2}$ in order to represent the strength and range of a typical atom-molecule potential~\cite{StoneBook13}.
Furthermore we set $a_\text{bb}=3.3\,(m B)^{-1/2}$, which approximately reproduces the speed of sound in superfluid $^4\text{He}$  for a molecule with $B=2\pi \times 1~\text{GHz}$ \cite{DonnellyHe98}.

The long-lived angulon states  correspond to the sharp features (dark shade) in Fig.~\ref{Stark}(a), and we find that their energy decreases with growing density. However, in addition to a uniform shift, also present in polarons~\cite{LandauPolaron,LandauPekarJETP48,AppelPolarons,Devreese13,Rath2013,Shashi2014,Shchadilova2014,Grusdt2014}, in the angulon problem the anisotropic molecule-boson interaction leads to an additional shift, whose magnitude  depends on $L$. Such a shift has been observed in spectra of molecules embedded in superfluid helium nanodroplets~\cite{ToenniesAngChem04, ZillichPRL04, ZillichPRB04, Babichenko99}, however it has never been derived from a general microscopic theory. By analogy with quantum electrodynamics (QED) \cite{Karshenboim2005}, this effect can be understood as ``rotational Lamb shift'' (RLS).  Fig.~\ref{Stark}(b) illustrates the RLS, defined as $\tilde{\Delta}^\text{RLS}_L = (E_L - E_0)/B - L(L+1)$, for a few lowest angular momentum states. The main feature is the change of sign of the RLS at the transition from the perturbative to the non-perturbative regime. As detailed below, this effect is closely related to the development of the many-body-induced fine structure of the second kind.  We find that the RLS is quite sensitive to the  anisotropy of the two-body potentials and therefore can be used as an accurate spectroscopic tool to  measure the microscopic parameters of the system.

 Even at low densities the magnitude of the RLS is sufficiently large to be measured in modern experiments on ultracold quantum gases.  For example, for a ground-state molecule with   $B= 2\pi\times 1$~GHz, trapped in a $^{23}\text{Na}$ BEC of density $10^{15}$~cm$^{-3}$ with $a_\text{bb}=60~\text{a.u.}$ \cite{Chin2010} we obtain a Lamb shift on the order of $10$~Hz. This prediction assumes molecule-atom interactions characterized by an $s$-wave scattering length $a=-100$~a.u., as well as an anisotropic $\lambda=1$ potential of comparable strength and finite range $r_1=100$~a.u. For ultracold molecules in highly-excited vibrational states with $B= 2\pi\times 1$~MHz  we find an increase of the Lamb shift by an order of magnitude. 

In addition to the RLS, a rich fine structure of the angulon excitation spectrum unfolds with increasing interaction strength $\tilde n$. In spirit of molecular spectroscopy~\cite{LevebvreBrionField2}, we  classify  the angulon states according to $L_{j, \Lambda}$, where in addition to the good quantum number, $L$, we introduced the ``approximate quantum numbers'', $j$ and $\Lambda$, indicating the dominant angular momentum contributions from the molecule and bosons, respectively. 
Starting from   weak interactions, the quasi-free molecular states, $L_{L, 0}$,  become dressed by bosonic excitations, which leads to a negative energy shift.  While the ground angulon state, $0_{00}^-$, is infinitely long lived, any of the $L>0$ states has a finite lifetime due to the weak coupling to phonons with $\lambda>0$. For states with $Z_L\approx 1$ the lifetime $\tau$ is given by Eq.~\eqref{ImSigma} as $\tau^{-1}= \text{Im} \Sigma^\text{ret}_L(E_L)$. The $\theta$-function of Eq.~\eqref{ImSigma}  determines which decay channels are open at a given  energy, while the Clebsch-Gordan coefficient gives the branching ratios between them.

As interactions increase with $\tilde n$, two distinct features appear in the angulon spectrum which we refer to as Many-Body-Induced Fine Structure (MBIFS) of the first and second kind. The MBIFS of the first kind is related to a new angulon state emerging  on top of the phonon continuum at positive energies when the interaction energy suffices to probe the finite range of the isotropic part of the interaction, $E\sim 1/(2 m r_0^2)$.  As a result, the state splits into two angulon branches, $L_{L,0} \to \{L_{L, 0}^-, L_{L, 0}^+\}$, as shown in Fig.~\ref{Stark}(c,d). An analysis of the  self-energy in Eq.~\eqref{Selfenergy} shows that this effect is solely due to the isotropic part of the finite-range potential. The appearance of the $L_{L, 0}^+$ state can be understood as a resonance in the many-body spectrum emerging due to coupling between the molecule and phonon states outside of the scattering continuum. Within our model, the resulting excited state, $L_{L, 0}^+$, is stabilized since the scattering continuum is exponentially suppressed by the finite range interaction, $r_0$ (imposing a cutoff onto the bath spectral function, cf. the spin-boson model~\cite{LeggettRMP87}). A similar stabilization mechanism appears  e.g.\ for the repulsively bound atom pairs in optical lattices \cite{Winkler2006}.  Furthermore, the $L_{L, 0}^+$ state is reminiscent of the excited bound phonon-polaron state in the Holstein model \cite{Gogolin1982,BoncaPRB99}.  However, the formation of the upper-branch angulon $L_{L,0}^+$ competes with the coupling of the molecule to bosons with non-zero angular momentum.  `On mass shell' emission of such bosons out of the phonon vacuum, a process analogous to  spontaneous emission in  QED, renders the $L^+$ state metastable and ultimately leads to its complete decoherence at large values of $\tilde n$.

The phonon-mediated interactions between different rotor states, $j$,  originate from anisotropies of the two-body potentials.  Once  these couplings reach a critical value, another peculiar effect occurs, which we refer to as MBIFS of the second kind.  Here, strong entanglement with the bath leads to the  disintegration of the lower-branch angulon, $L^-_{L,0}$, into the phonon continuum belonging to the rotor state with $j=L-1$.  The $L^-$ state then reappears at lower energy separated by a gap which scales with the anisotropic part of the potential, $u_1$. An analysis of the occupation numbers $\beta^{L}_{k\lambda j}$  of Eq.~\eqref{VarFunc}  reveals that at the transition, approximately one quantum of angular momentum is transferred from the rotor to the many-body bath, $L^-_{L,0} \to L_{L-1, 1}^-$.  Within the single-phonon excitation approach of Eq.~(\ref{VarFunc}), the reemerging angulon has an infinite lifetime as soon as it lies energetically below the scattering continuum of the $L-1$ state as can be seen analytically from Eq.~\eqref{ImSigma}. The stability threshold for $1_{0,1}^-$ state is shown in Fig.~\ref{Stark}(a) by the horizontal dashed line.   Both the MBIFS of the first and second kind can be probed spectroscopically at the typical parameters of molecules in helium nanodroplets~\cite{ToenniesAngChem04}; their  spectroscopic signatures are shown in Fig.~\ref{Stark}(d).

In summary, we developed a quasiparticle-based approach to the redistribution of angular momentum between a rotating impurity and a bosonic bath.   Our theory sheds light on the properties of molecules trapped in superfluid helium droplets from a general many-body perspective, and predicts a rich rotational fine-structure arising due to the interaction with a many-particle environment. Accounting for the predicted effects might be key to the understanding of recent experimental data on non-adiabatic molecular excitations~\cite{PentlehnerPRL13}, which so far lacks even a qualitative explanation. In a broader perspective, the angulon can serve as a building block for the description of few-, and many-molecule processes in the presence of an environment, such as reactivity and molecular collisions, and provide an efficient description of quantum rotor models \cite{SachdevBook} coupled to an external bath. The presented theory can be extended to account for complex molecular potentials and vibrations~\cite{StoneBook13}, external fields~\cite{LemKreDoyKais13}, the physics of rotons and maxons \cite{DonnellyHe98},  multi-phonon excitations, as well as phonon interactions and decay \cite{Beliaev1958,Giorgini2000} which is expected to further enrich the observed phenomena.

We are grateful to Eugene Demler, Sarang Gopalakrishnan,  Michael Knap, Vasili Kharchenko, 
Markus Oberthaler, Seth Rittenhouse, Robert Seiringer, and Zhenhua Yu for fruitful discussions. R.S. is grateful for the hospitality during his stay at the Institute for Advanced Study at Tsinghua University. The work was supported by the NSF through a grant for the Institute for Theoretical Atomic, Molecular, and Optical Physics at Harvard University and Smithsonian Astrophysical Observatory.


\begin{thebibliography}{46}%
\makeatletter
\providecommand \@ifxundefined [1]{%
 \@ifx{#1\undefined}
}%
\providecommand \@ifnum [1]{%
 \ifnum #1\expandafter \@firstoftwo
 \else \expandafter \@secondoftwo
 \fi
}%
\providecommand \@ifx [1]{%
 \ifx #1\expandafter \@firstoftwo
 \else \expandafter \@secondoftwo
 \fi
}%
\providecommand \natexlab [1]{#1}%
\providecommand \enquote  [1]{``#1''}%
\providecommand \bibnamefont  [1]{#1}%
\providecommand \bibfnamefont [1]{#1}%
\providecommand \citenamefont [1]{#1}%
\providecommand \href@noop [0]{\@secondoftwo}%
\providecommand \href [0]{\begingroup \@sanitize@url \@href}%
\providecommand \@href[1]{\@@startlink{#1}\@@href}%
\providecommand \@@href[1]{\endgroup#1\@@endlink}%
\providecommand \@sanitize@url [0]{\catcode `\\12\catcode `\$12\catcode
  `\&12\catcode `\#12\catcode `\^12\catcode `\_12\catcode `\%12\relax}%
\providecommand \@@startlink[1]{}%
\providecommand \@@endlink[0]{}%
\providecommand \url  [0]{\begingroup\@sanitize@url \@url }%
\providecommand \@url [1]{\endgroup\@href {#1}{\urlprefix }}%
\providecommand \urlprefix  [0]{URL }%
\providecommand \Eprint [0]{\href }%
\providecommand \doibase [0]{http://dx.doi.org/}%
\providecommand \selectlanguage [0]{\@gobble}%
\providecommand \bibinfo  [0]{\@secondoftwo}%
\providecommand \bibfield  [0]{\@secondoftwo}%
\providecommand \translation [1]{[#1]}%
\providecommand \BibitemOpen [0]{}%
\providecommand \bibitemStop [0]{}%
\providecommand \bibitemNoStop [0]{.\EOS\space}%
\providecommand \EOS [0]{\spacefactor3000\relax}%
\providecommand \BibitemShut  [1]{\csname bibitem#1\endcsname}%
\let\auto@bib@innerbib\@empty
\bibitem [{\citenamefont {Walecka}(2004)}]{WaleckaNuclearBook}%
  \BibitemOpen
  \bibfield  {author} {\bibinfo {author} {\bibfnamefont {J.~D.}\ \bibnamefont
  {Walecka}},\ }\href@noop {} {\emph {\bibinfo {title} {Theoretical Nuclear and
  Subnuclear Physics}}},\ \bibinfo {edition} {2nd}\ ed.\ (\bibinfo  {publisher}
  {World Scientific, Imperial College Press},\ \bibinfo {year}
  {2004})\BibitemShut {NoStop}%
\bibitem [{\citenamefont {Derevianko}\ and\ \citenamefont
  {Katori}(2011)}]{DereviankoRMP11}%
  \BibitemOpen
  \bibfield  {author} {\bibinfo {author} {\bibfnamefont {A.}~\bibnamefont
  {Derevianko}}\ and\ \bibinfo {author} {\bibfnamefont {H.}~\bibnamefont
  {Katori}},\ }\href@noop {} {\bibfield  {journal} {\bibinfo  {journal} {Rev.
  Mod. Phys.}\ }\textbf {\bibinfo {volume} {83}},\ \bibinfo {pages} {331}
  (\bibinfo {year} {2011})}\BibitemShut {NoStop}%
\bibitem [{\citenamefont {Maze}\ \emph {et~al.}(2011)\citenamefont {Maze},
  \citenamefont {Gali}, \citenamefont {Togan}, \citenamefont {Chu},
  \citenamefont {Trifonov}, \citenamefont {Kaxiras},\ and\ \citenamefont
  {Lukin}}]{MazeNJP11}%
  \BibitemOpen
  \bibfield  {author} {\bibinfo {author} {\bibfnamefont {J.}~\bibnamefont
  {Maze}}, \bibinfo {author} {\bibfnamefont {A.}~\bibnamefont {Gali}}, \bibinfo
  {author} {\bibfnamefont {E.}~\bibnamefont {Togan}}, \bibinfo {author}
  {\bibfnamefont {Y.}~\bibnamefont {Chu}}, \bibinfo {author} {\bibfnamefont
  {A.}~\bibnamefont {Trifonov}}, \bibinfo {author} {\bibfnamefont
  {E.}~\bibnamefont {Kaxiras}}, \ and\ \bibinfo {author} {\bibfnamefont
  {M.~D.}\ \bibnamefont {Lukin}},\ }\href@noop {} {\bibfield  {journal}
  {\bibinfo  {journal} {New Journal of Physics}\ }\textbf {\bibinfo {volume}
  {13}},\ \bibinfo {pages} {025025} (\bibinfo {year} {2011})}\BibitemShut
  {NoStop}%
\bibitem [{\citenamefont {Wigner}\ and\ \citenamefont
  {Griffin}(1959)}]{WignerGroupTheory}%
  \BibitemOpen
  \bibfield  {author} {\bibinfo {author} {\bibfnamefont {E.~P.}\ \bibnamefont
  {Wigner}}\ and\ \bibinfo {author} {\bibfnamefont {J.~J.}\ \bibnamefont
  {Griffin}},\ }\href@noop {} {\emph {\bibinfo {title} {Group Theory and Its
  Application to the Quantum Mechanics of Atomic Spectra}}}\ (\bibinfo
  {publisher} {Academic Press},\ \bibinfo {year} {1959})\BibitemShut {NoStop}%
\bibitem [{\citenamefont {Racah}(1942{\natexlab{a}})}]{RacahComplSpec1}%
  \BibitemOpen
  \bibfield  {author} {\bibinfo {author} {\bibfnamefont {G.}~\bibnamefont
  {Racah}},\ }\href@noop {} {\bibfield  {journal} {\bibinfo  {journal} {Phys.
  Rev.}\ }\textbf {\bibinfo {volume} {61}},\ \bibinfo {pages} {186} (\bibinfo
  {year} {1942}{\natexlab{a}})}\BibitemShut {NoStop}%
\bibitem [{\citenamefont {Racah}(1942{\natexlab{b}})}]{RacahComplSpec2}%
  \BibitemOpen
  \bibfield  {author} {\bibinfo {author} {\bibfnamefont {G.}~\bibnamefont
  {Racah}},\ }\href@noop {} {\bibfield  {journal} {\bibinfo  {journal} {Phys.
  Rev.}\ }\textbf {\bibinfo {volume} {62}},\ \bibinfo {pages} {438} (\bibinfo
  {year} {1942}{\natexlab{b}})}\BibitemShut {NoStop}%
\bibitem [{\citenamefont {Racah}(1943)}]{RacahComplSpec3}%
  \BibitemOpen
  \bibfield  {author} {\bibinfo {author} {\bibfnamefont {G.}~\bibnamefont
  {Racah}},\ }\href@noop {} {\bibfield  {journal} {\bibinfo  {journal} {Phys.
  Rev.}\ }\textbf {\bibinfo {volume} {63}},\ \bibinfo {pages} {367} (\bibinfo
  {year} {1943})}\BibitemShut {NoStop}%
\bibitem [{\citenamefont {Varshalovich}\ \emph {et~al.}(1988)\citenamefont
  {Varshalovich}, \citenamefont {Moskalev},\ and\ \citenamefont
  {Khersonski}}]{VarshalovichAngMom}%
  \BibitemOpen
  \bibfield  {author} {\bibinfo {author} {\bibfnamefont {D.~A.}\ \bibnamefont
  {Varshalovich}}, \bibinfo {author} {\bibfnamefont {A.~N.}\ \bibnamefont
  {Moskalev}}, \ and\ \bibinfo {author} {\bibfnamefont {V.~K.}\ \bibnamefont
  {Khersonski}},\ }\href@noop {} {\emph {\bibinfo {title} {Quantum theory of
  angular momentum}}}\ (\bibinfo  {publisher} {World Scientific Publications,
  Singapore and Teaneck, N.J.},\ \bibinfo {year} {1988})\BibitemShut {NoStop}%
\bibitem [{\citenamefont {Zare}(1988)}]{ZareAngMom}%
  \BibitemOpen
  \bibfield  {author} {\bibinfo {author} {\bibfnamefont {R.~N.}\ \bibnamefont
  {Zare}},\ }\href@noop {} {\emph {\bibinfo {title} {Angular momentum:
  Understanding spatial aspects in chemistry and physics}}}\ (\bibinfo
  {publisher} {Wiley, New York},\ \bibinfo {year} {1988})\BibitemShut {NoStop}%
\bibitem [{\citenamefont {Abrikosov}\ \emph {et~al.}(1963)\citenamefont
  {Abrikosov}, \citenamefont {Gorkov},\ and\ \citenamefont
  {Dzyaloshinski}}]{AbrikosovBook}%
  \BibitemOpen
  \bibfield  {author} {\bibinfo {author} {\bibfnamefont {A.~A.}\ \bibnamefont
  {Abrikosov}}, \bibinfo {author} {\bibfnamefont {L.~P.}\ \bibnamefont
  {Gorkov}}, \ and\ \bibinfo {author} {\bibfnamefont {I.~E.}\ \bibnamefont
  {Dzyaloshinski}},\ }\href@noop {} {\emph {\bibinfo {title} {Methods of
  Quantum Field Theory in Statistical Physics}}}\ (\bibinfo  {publisher}
  {Prentice-Hall, NJ},\ \bibinfo {year} {1963})\BibitemShut {NoStop}%
\bibitem [{\citenamefont {Weiss}(2012)}]{WeissBook}%
  \BibitemOpen
  \bibfield  {author} {\bibinfo {author} {\bibfnamefont {U.}~\bibnamefont
  {Weiss}},\ }\href@noop {} {\emph {\bibinfo {title} {Quantum Dissipative
  Systems}}},\ \bibinfo {edition} {4th}\ ed.\ (\bibinfo  {publisher} {World
  Scientific},\ \bibinfo {year} {2012})\BibitemShut {NoStop}%
\bibitem [{\citenamefont {Breuer}\ and\ \citenamefont
  {Petruccione}(2002)}]{Breuer2002}%
  \BibitemOpen
  \bibfield  {author} {\bibinfo {author} {\bibfnamefont {H.-P.}\ \bibnamefont
  {Breuer}}\ and\ \bibinfo {author} {\bibfnamefont {F.}~\bibnamefont
  {Petruccione}},\ }\href@noop {} {\emph {\bibinfo {title} {The {T}heory of
  {O}pen {Q}uantum {S}ystems}}}\ (\bibinfo  {publisher} {Oxford University
  Press},\ \bibinfo {address} {Oxford},\ \bibinfo {year} {2002})\BibitemShut
  {NoStop}%
\bibitem [{\citenamefont {Landau}(1933)}]{LandauPolaron}%
  \BibitemOpen
  \bibfield  {author} {\bibinfo {author} {\bibfnamefont {L.~D.}\ \bibnamefont
  {Landau}},\ }\href@noop {} {\bibfield  {journal} {\bibinfo  {journal} {Phys.
  Z. Sowjetunion}\ }\textbf {\bibinfo {volume} {3}},\ \bibinfo {pages} {664}
  (\bibinfo {year} {1933})}\BibitemShut {NoStop}%
\bibitem [{\citenamefont {Landau}\ and\ \citenamefont
  {Pekar}(1948)}]{LandauPekarJETP48}%
  \BibitemOpen
  \bibfield  {author} {\bibinfo {author} {\bibfnamefont {L.~D.}\ \bibnamefont
  {Landau}}\ and\ \bibinfo {author} {\bibfnamefont {S.~I.}\ \bibnamefont
  {Pekar}},\ }\href@noop {} {\bibfield  {journal} {\bibinfo  {journal} {Zh.
  Eksp. i Theor. Fiz.}\ }\textbf {\bibinfo {volume} {18}},\ \bibinfo {pages}
  {419} (\bibinfo {year} {1948})}\BibitemShut {NoStop}%
\bibitem [{\citenamefont {Appel}(1968)}]{AppelPolarons}%
  \BibitemOpen
  \bibfield  {author} {\bibinfo {author} {\bibfnamefont {J.}~\bibnamefont
  {Appel}},\ }in\ \href@noop {} {\emph {\bibinfo {booktitle} {Solid State
  Physics}}},\ Vol.~\bibinfo {volume} {21},\ \bibinfo {editor} {edited by\
  \bibinfo {editor} {\bibfnamefont {H.}~\bibnamefont {Ehrenreich}}, \bibinfo
  {editor} {\bibfnamefont {F.}~\bibnamefont {Seitz}}, \ and\ \bibinfo {editor}
  {\bibfnamefont {D.}~\bibnamefont {Turnbull}}}\ (\bibinfo  {publisher}
  {Academic, NY},\ \bibinfo {year} {1968})\BibitemShut {NoStop}%
\bibitem [{\citenamefont {Devreese}(2013)}]{Devreese13}%
  \BibitemOpen
  \bibfield  {author} {\bibinfo {author} {\bibfnamefont {J.~T.}\ \bibnamefont
  {Devreese}},\ }\href@noop {} {\bibfield  {journal} {\bibinfo  {journal}
  {arXiv:1012.4576}\ } (\bibinfo {year} {2013})}\BibitemShut {NoStop}%
\bibitem [{\citenamefont {Leggett}\ \emph {et~al.}(1987)\citenamefont
  {Leggett}, \citenamefont {Chakravarty}, \citenamefont {Dorsey}, \citenamefont
  {Fisher}, \citenamefont {Garg},\ and\ \citenamefont
  {Zwerger}}]{LeggettRMP87}%
  \BibitemOpen
  \bibfield  {author} {\bibinfo {author} {\bibfnamefont {A.}~\bibnamefont
  {Leggett}}, \bibinfo {author} {\bibfnamefont {S.}~\bibnamefont
  {Chakravarty}}, \bibinfo {author} {\bibfnamefont {A.~T.}\ \bibnamefont
  {Dorsey}}, \bibinfo {author} {\bibfnamefont {M.~P.~A.}\ \bibnamefont
  {Fisher}}, \bibinfo {author} {\bibfnamefont {A.}~\bibnamefont {Garg}}, \ and\
  \bibinfo {author} {\bibfnamefont {W.}~\bibnamefont {Zwerger}},\ }\href@noop
  {} {\bibfield  {journal} {\bibinfo  {journal} {Rev. Mod. Phys.}\ }\textbf
  {\bibinfo {volume} {59}},\ \bibinfo {pages} {1} (\bibinfo {year}
  {1987})}\BibitemShut {NoStop}%
\bibitem [{\citenamefont {Toennies}\ and\ \citenamefont
  {Vilesov}(2004)}]{ToenniesAngChem04}%
  \BibitemOpen
  \bibfield  {author} {\bibinfo {author} {\bibfnamefont {J.~P.}\ \bibnamefont
  {Toennies}}\ and\ \bibinfo {author} {\bibfnamefont {A.~F.}\ \bibnamefont
  {Vilesov}},\ }\href@noop {} {\bibfield  {journal} {\bibinfo  {journal}
  {Angewandte Chemie International Edition}\ }\textbf {\bibinfo {volume}
  {43}},\ \bibinfo {pages} {2622} (\bibinfo {year} {2004})}\BibitemShut
  {NoStop}%
\bibitem [{\citenamefont {Mahan}(1990)}]{Mahan90}%
  \BibitemOpen
  \bibfield  {author} {\bibinfo {author} {\bibfnamefont {G.~D.}\ \bibnamefont
  {Mahan}},\ }\href@noop {} {\emph {\bibinfo {title} {Many-particle
  physics}}},\ Physics of solids and liquids\ (\bibinfo  {publisher} {Plenum},\
  \bibinfo {address} {New York, NY},\ \bibinfo {year} {1990})\BibitemShut
  {NoStop}%
\bibitem [{\citenamefont {Gardiner}\ and\ \citenamefont
  {Zoller}(2000)}]{GardinerZollerBook}%
  \BibitemOpen
  \bibfield  {author} {\bibinfo {author} {\bibfnamefont {C.~W.}\ \bibnamefont
  {Gardiner}}\ and\ \bibinfo {author} {\bibfnamefont {P.}~\bibnamefont
  {Zoller}},\ }\href@noop {} {\emph {\bibinfo {title} {Quantum Noise}}}\
  (\bibinfo  {publisher} {Springer},\ \bibinfo {year} {2000})\BibitemShut
  {NoStop}%
\bibitem [{\citenamefont {Pitaevskii}\ and\ \citenamefont
  {Stringari}(2003)}]{Pitaevskii2003}%
  \BibitemOpen
  \bibfield  {author} {\bibinfo {author} {\bibfnamefont {L.~P.}\ \bibnamefont
  {Pitaevskii}}\ and\ \bibinfo {author} {\bibfnamefont {S.}~\bibnamefont
  {Stringari}},\ }\href@noop {} {\emph {\bibinfo {title} {Bose-Einstein
  Condensation}}}\ (\bibinfo  {publisher} {Oxford: Clarendon},\ \bibinfo {year}
  {2003})\BibitemShut {NoStop}%
\bibitem [{\citenamefont {Lefebvre-Brion}\ and\ \citenamefont
  {Field}(2004)}]{LevebvreBrionField2}%
  \BibitemOpen
  \bibfield  {author} {\bibinfo {author} {\bibfnamefont {H.}~\bibnamefont
  {Lefebvre-Brion}}\ and\ \bibinfo {author} {\bibfnamefont {R.~W.}\
  \bibnamefont {Field}},\ }\href@noop {} {\emph {\bibinfo {title} {The Spectra
  and Dynamics of Diatomic Molecules}}}\ (\bibinfo  {publisher} {Elsevier, New
  York},\ \bibinfo {year} {2004})\BibitemShut {NoStop}%
\bibitem [{\citenamefont {Lemeshko}\ \emph {et~al.}(2013)\citenamefont
  {Lemeshko}, \citenamefont {Krems}, \citenamefont {Doyle},\ and\ \citenamefont
  {Kais}}]{LemKreDoyKais13}%
  \BibitemOpen
  \bibfield  {author} {\bibinfo {author} {\bibfnamefont {M.}~\bibnamefont
  {Lemeshko}}, \bibinfo {author} {\bibfnamefont {R.}~\bibnamefont {Krems}},
  \bibinfo {author} {\bibfnamefont {J.}~\bibnamefont {Doyle}}, \ and\ \bibinfo
  {author} {\bibfnamefont {S.}~\bibnamefont {Kais}},\ }\href@noop {} {\bibfield
   {journal} {\bibinfo  {journal} {Mol. Phys.}\ }\textbf {\bibinfo {volume}
  {111}},\ \bibinfo {pages} {1648} (\bibinfo {year} {2013})}\BibitemShut
  {NoStop}%
\bibitem [{sup()}]{sup}%
  \BibitemOpen
  \href@noop {} {\emph {\bibinfo {title} {\textnormal{See the supplemental
  material}}}}\BibitemShut {NoStop}%
\bibitem [{\citenamefont {Wang}\ \emph {et~al.}(2005)\citenamefont {Wang},
  \citenamefont {Lukin},\ and\ \citenamefont {Demler}}]{Wang2005}%
  \BibitemOpen
  \bibfield  {author} {\bibinfo {author} {\bibfnamefont {D.-W.}\ \bibnamefont
  {Wang}}, \bibinfo {author} {\bibfnamefont {M.~D.}\ \bibnamefont {Lukin}}, \
  and\ \bibinfo {author} {\bibfnamefont {E.}~\bibnamefont {Demler}},\ }\href
  {\doibase 10.1103/PhysRevA.72.051604} {\bibfield  {journal} {\bibinfo
  {journal} {Phys. Rev. A}\ }\textbf {\bibinfo {volume} {72}},\ \bibinfo
  {pages} {051604} (\bibinfo {year} {2005})}\BibitemShut {NoStop}%
\bibitem [{\citenamefont {Wang}(2006)}]{Wang2006}%
  \BibitemOpen
  \bibfield  {author} {\bibinfo {author} {\bibfnamefont {D.-W.}\ \bibnamefont
  {Wang}},\ }\href {\doibase 10.1103/PhysRevLett.96.140404} {\bibfield
  {journal} {\bibinfo  {journal} {Phys. Rev. Lett.}\ }\textbf {\bibinfo
  {volume} {96}},\ \bibinfo {pages} {140404} (\bibinfo {year}
  {2006})}\BibitemShut {NoStop}%
\bibitem [{\citenamefont {Tempere}\ \emph {et~al.}(2009)\citenamefont
  {Tempere}, \citenamefont {Casteels}, \citenamefont {Oberthaler},
  \citenamefont {Knoop}, \citenamefont {Timmermans},\ and\ \citenamefont
  {Devreese}}]{TemperePRB09}%
  \BibitemOpen
  \bibfield  {author} {\bibinfo {author} {\bibfnamefont {J.}~\bibnamefont
  {Tempere}}, \bibinfo {author} {\bibfnamefont {W.}~\bibnamefont {Casteels}},
  \bibinfo {author} {\bibfnamefont {M.}~\bibnamefont {Oberthaler}}, \bibinfo
  {author} {\bibfnamefont {S.}~\bibnamefont {Knoop}}, \bibinfo {author}
  {\bibfnamefont {E.}~\bibnamefont {Timmermans}}, \ and\ \bibinfo {author}
  {\bibfnamefont {J.}~\bibnamefont {Devreese}},\ }\href@noop {} {\bibfield
  {journal} {\bibinfo  {journal} {Physical Review B}\ }\textbf {\bibinfo
  {volume} {80}},\ \bibinfo {pages} {184504} (\bibinfo {year}
  {2009})}\BibitemShut {NoStop}%
\bibitem [{\citenamefont {Hartmann}\ \emph {et~al.}(2002)\citenamefont
  {Hartmann}, \citenamefont {Lindinger}, \citenamefont {Toennies},\ and\
  \citenamefont {Vilesov}}]{Hartmann2002}%
  \BibitemOpen
  \bibfield  {author} {\bibinfo {author} {\bibfnamefont {M.}~\bibnamefont
  {Hartmann}}, \bibinfo {author} {\bibfnamefont {A.}~\bibnamefont {Lindinger}},
  \bibinfo {author} {\bibfnamefont {J.~P.}\ \bibnamefont {Toennies}}, \ and\
  \bibinfo {author} {\bibfnamefont {A.~F.}\ \bibnamefont {Vilesov}},\
  }\href@noop {} {\bibfield  {journal} {\bibinfo  {journal} {Phys. Chem. Chem.
  Phys.}\ }\textbf {\bibinfo {volume} {4}},\ \bibinfo {pages} {4839} (\bibinfo
  {year} {2002})}\BibitemShut {NoStop}%
\bibitem [{\citenamefont {Stone}(2013)}]{StoneBook13}%
  \BibitemOpen
  \bibfield  {author} {\bibinfo {author} {\bibfnamefont {A.}~\bibnamefont
  {Stone}},\ }\href@noop {} {\emph {\bibinfo {title} {The Theory of
  Intermolecular Forces}}}\ (\bibinfo  {publisher} {Oxford University Press},\
  \bibinfo {year} {2013})\BibitemShut {NoStop}%
\bibitem [{\citenamefont {Donnelly}\ and\ \citenamefont
  {Barenghi}(1998)}]{DonnellyHe98}%
  \BibitemOpen
  \bibfield  {author} {\bibinfo {author} {\bibfnamefont {R.~J.}\ \bibnamefont
  {Donnelly}}\ and\ \bibinfo {author} {\bibfnamefont {C.~F.}\ \bibnamefont
  {Barenghi}},\ }\href@noop {} {\bibfield  {journal} {\bibinfo  {journal} {J.
  Phys. Chem. Ref. Data}\ }\textbf {\bibinfo {volume} {27}} (\bibinfo {year}
  {1998})}\BibitemShut {NoStop}%
\bibitem [{\citenamefont {Rath}\ and\ \citenamefont
  {Schmidt}(2013)}]{Rath2013}%
  \BibitemOpen
  \bibfield  {author} {\bibinfo {author} {\bibfnamefont {S.~P.}\ \bibnamefont
  {Rath}}\ and\ \bibinfo {author} {\bibfnamefont {R.}~\bibnamefont {Schmidt}},\
  }\href {\doibase 10.1103/PhysRevA.88.053632} {\bibfield  {journal} {\bibinfo
  {journal} {Phys. Rev. A}\ }\textbf {\bibinfo {volume} {88}},\ \bibinfo
  {pages} {053632} (\bibinfo {year} {2013})}\BibitemShut {NoStop}%
\bibitem [{\citenamefont {Shashi}\ \emph {et~al.}(2014)\citenamefont {Shashi},
  \citenamefont {Grusdt}, \citenamefont {Abanin},\ and\ \citenamefont
  {Demler}}]{Shashi2014}%
  \BibitemOpen
  \bibfield  {author} {\bibinfo {author} {\bibfnamefont {A.}~\bibnamefont
  {Shashi}}, \bibinfo {author} {\bibfnamefont {F.}~\bibnamefont {Grusdt}},
  \bibinfo {author} {\bibfnamefont {D.~A.}\ \bibnamefont {Abanin}}, \ and\
  \bibinfo {author} {\bibfnamefont {E.}~\bibnamefont {Demler}},\ }\href
  {\doibase 10.1103/PhysRevA.89.053617} {\bibfield  {journal} {\bibinfo
  {journal} {Phys. Rev. A}\ }\textbf {\bibinfo {volume} {89}},\ \bibinfo
  {pages} {053617} (\bibinfo {year} {2014})}\BibitemShut {NoStop}%
\bibitem [{\citenamefont {Shchadilova}\ \emph {et~al.}(2014)\citenamefont
  {Shchadilova}, \citenamefont {Grusdt}, \citenamefont {Rubtsov},\ and\
  \citenamefont {Demler}}]{Shchadilova2014}%
  \BibitemOpen
  \bibfield  {author} {\bibinfo {author} {\bibfnamefont {Y.~E.}\ \bibnamefont
  {Shchadilova}}, \bibinfo {author} {\bibfnamefont {F.}~\bibnamefont {Grusdt}},
  \bibinfo {author} {\bibfnamefont {A.~N.}\ \bibnamefont {Rubtsov}}, \ and\
  \bibinfo {author} {\bibfnamefont {E.}~\bibnamefont {Demler}},\ }\href@noop {}
  {\bibfield  {journal} {\bibinfo  {journal} {arXiv preprint arXiv:1410.5691}\
  } (\bibinfo {year} {2014})}\BibitemShut {NoStop}%
\bibitem [{\citenamefont {Grusdt}\ \emph {et~al.}(2014)\citenamefont {Grusdt},
  \citenamefont {Shchadilova}, \citenamefont {Rubtsov},\ and\ \citenamefont
  {Demler}}]{Grusdt2014}%
  \BibitemOpen
  \bibfield  {author} {\bibinfo {author} {\bibfnamefont {F.}~\bibnamefont
  {Grusdt}}, \bibinfo {author} {\bibfnamefont {Y.}~\bibnamefont {Shchadilova}},
  \bibinfo {author} {\bibfnamefont {A.}~\bibnamefont {Rubtsov}}, \ and\
  \bibinfo {author} {\bibfnamefont {E.}~\bibnamefont {Demler}},\ }\href@noop {}
  {\bibfield  {journal} {\bibinfo  {journal} {arXiv preprint arXiv:1410.2203}\
  } (\bibinfo {year} {2014})}\BibitemShut {NoStop}%
\bibitem [{\citenamefont {Zillich}\ \emph {et~al.}(2004)\citenamefont
  {Zillich}, \citenamefont {Kwon},\ and\ \citenamefont
  {Whaley}}]{ZillichPRL04}%
  \BibitemOpen
  \bibfield  {author} {\bibinfo {author} {\bibfnamefont {R.~E.}\ \bibnamefont
  {Zillich}}, \bibinfo {author} {\bibfnamefont {Y.}~\bibnamefont {Kwon}}, \
  and\ \bibinfo {author} {\bibfnamefont {K.~B.}\ \bibnamefont {Whaley}},\
  }\href@noop {} {\bibfield  {journal} {\bibinfo  {journal} {Phys. Rev. Lett.}\
  }\textbf {\bibinfo {volume} {93}},\ \bibinfo {pages} {250401} (\bibinfo
  {year} {2004})}\BibitemShut {NoStop}%
\bibitem [{\citenamefont {Zillich}\ and\ \citenamefont
  {Whaley}(2004)}]{ZillichPRB04}%
  \BibitemOpen
  \bibfield  {author} {\bibinfo {author} {\bibfnamefont {R.~E.}\ \bibnamefont
  {Zillich}}\ and\ \bibinfo {author} {\bibfnamefont {K.~B.}\ \bibnamefont
  {Whaley}},\ }\href@noop {} {\bibfield  {journal} {\bibinfo  {journal} {Phys.
  Rev. B}\ }\textbf {\bibinfo {volume} {69}},\ \bibinfo {pages} {104517}
  (\bibinfo {year} {2004})}\BibitemShut {NoStop}%
\bibitem [{\citenamefont {Babichenko}\ and\ \citenamefont
  {Kagan}(1999)}]{Babichenko99}%
  \BibitemOpen
  \bibfield  {author} {\bibinfo {author} {\bibfnamefont {V.~S.}\ \bibnamefont
  {Babichenko}}\ and\ \bibinfo {author} {\bibfnamefont {Y.}~\bibnamefont
  {Kagan}},\ }\href@noop {} {\bibfield  {journal} {\bibinfo  {journal} {Phys.
  Rev. Lett.}\ }\textbf {\bibinfo {volume} {83}} (\bibinfo {year}
  {1999})}\BibitemShut {NoStop}%
\bibitem [{\citenamefont {Karshenboim}(2005)}]{Karshenboim2005}%
  \BibitemOpen
  \bibfield  {author} {\bibinfo {author} {\bibfnamefont {S.~G.}\ \bibnamefont
  {Karshenboim}},\ }\href@noop {} {\bibfield  {journal} {\bibinfo  {journal}
  {Phys. Rep.}\ }\textbf {\bibinfo {volume} {422}},\ \bibinfo {pages} {1}
  (\bibinfo {year} {2005})}\BibitemShut {NoStop}%
\bibitem [{\citenamefont {Chin}\ \emph {et~al.}(2010)\citenamefont {Chin},
  \citenamefont {Grimm}, \citenamefont {Julienne},\ and\ \citenamefont
  {Tiesinga}}]{Chin2010}%
  \BibitemOpen
  \bibfield  {author} {\bibinfo {author} {\bibfnamefont {C.}~\bibnamefont
  {Chin}}, \bibinfo {author} {\bibfnamefont {R.}~\bibnamefont {Grimm}},
  \bibinfo {author} {\bibfnamefont {P.}~\bibnamefont {Julienne}}, \ and\
  \bibinfo {author} {\bibfnamefont {E.}~\bibnamefont {Tiesinga}},\ }\href
  {\doibase 10.1103/RevModPhys.82.1225} {\bibfield  {journal} {\bibinfo
  {journal} {Rev. Mod. Phys.}\ }\textbf {\bibinfo {volume} {82}},\ \bibinfo
  {pages} {1225} (\bibinfo {year} {2010})}\BibitemShut {NoStop}%
\bibitem [{\citenamefont {Winkler}\ \emph {et~al.}(2006)\citenamefont
  {Winkler}, \citenamefont {Thalhammer}, \citenamefont {Lang}, \citenamefont
  {Grimm}, \citenamefont {Denschlag}, \citenamefont {Daley}, \citenamefont
  {Kantian}, \citenamefont {B{\"u}chler},\ and\ \citenamefont
  {Zoller}}]{Winkler2006}%
  \BibitemOpen
  \bibfield  {author} {\bibinfo {author} {\bibfnamefont {K.}~\bibnamefont
  {Winkler}}, \bibinfo {author} {\bibfnamefont {G.}~\bibnamefont {Thalhammer}},
  \bibinfo {author} {\bibfnamefont {F.}~\bibnamefont {Lang}}, \bibinfo {author}
  {\bibfnamefont {R.}~\bibnamefont {Grimm}}, \bibinfo {author} {\bibfnamefont
  {J.~H.}\ \bibnamefont {Denschlag}}, \bibinfo {author} {\bibfnamefont
  {A.}~\bibnamefont {Daley}}, \bibinfo {author} {\bibfnamefont
  {A.}~\bibnamefont {Kantian}}, \bibinfo {author} {\bibfnamefont
  {H.}~\bibnamefont {B{\"u}chler}}, \ and\ \bibinfo {author} {\bibfnamefont
  {P.}~\bibnamefont {Zoller}},\ }\href@noop {} {\bibfield  {journal} {\bibinfo
  {journal} {Nature}\ }\textbf {\bibinfo {volume} {441}},\ \bibinfo {pages}
  {853} (\bibinfo {year} {2006})}\BibitemShut {NoStop}%
\bibitem [{\citenamefont {Gogolin}(1982)}]{Gogolin1982}%
  \BibitemOpen
  \bibfield  {author} {\bibinfo {author} {\bibfnamefont {A.}~\bibnamefont
  {Gogolin}},\ }\href@noop {} {\bibfield  {journal} {\bibinfo  {journal} {Phys.
  Stat. Solidi B}\ }\textbf {\bibinfo {volume} {109}},\ \bibinfo {pages} {95}
  (\bibinfo {year} {1982})}\BibitemShut {NoStop}%
\bibitem [{\citenamefont {Bon\v{c}a}\ \emph {et~al.}(1999)\citenamefont
  {Bon\v{c}a}, \citenamefont {Trugman},\ and\ \citenamefont
  {Batisti\'{c}}}]{BoncaPRB99}%
  \BibitemOpen
  \bibfield  {author} {\bibinfo {author} {\bibfnamefont {J.}~\bibnamefont
  {Bon\v{c}a}}, \bibinfo {author} {\bibfnamefont {S.~A.}\ \bibnamefont
  {Trugman}}, \ and\ \bibinfo {author} {\bibfnamefont {I.}~\bibnamefont
  {Batisti\'{c}}},\ }\href@noop {} {\bibfield  {journal} {\bibinfo  {journal}
  {Phys. Rev. B}\ }\textbf {\bibinfo {volume} {60}},\ \bibinfo {pages} {1633}
  (\bibinfo {year} {1999})}\BibitemShut {NoStop}%
\bibitem [{\citenamefont {Pentlehner}\ \emph {et~al.}(2013)\citenamefont
  {Pentlehner}, \citenamefont {Nielsen}, \citenamefont {Slenczka},
  \citenamefont {M{\o}lmer},\ and\ \citenamefont
  {Stapelfeldt}}]{PentlehnerPRL13}%
  \BibitemOpen
  \bibfield  {author} {\bibinfo {author} {\bibfnamefont {D.}~\bibnamefont
  {Pentlehner}}, \bibinfo {author} {\bibfnamefont {J.~H.}\ \bibnamefont
  {Nielsen}}, \bibinfo {author} {\bibfnamefont {A.}~\bibnamefont {Slenczka}},
  \bibinfo {author} {\bibfnamefont {K.}~\bibnamefont {M{\o}lmer}}, \ and\
  \bibinfo {author} {\bibfnamefont {H.}~\bibnamefont {Stapelfeldt}},\
  }\href@noop {} {\bibfield  {journal} {\bibinfo  {journal} {Phys. Rev. Lett.}\
  }\textbf {\bibinfo {volume} {110}},\ \bibinfo {pages} {093002} (\bibinfo
  {year} {2013})}\BibitemShut {NoStop}%
\bibitem [{\citenamefont {Sachdev}(2011)}]{SachdevBook}%
  \BibitemOpen
  \bibfield  {author} {\bibinfo {author} {\bibfnamefont {S.}~\bibnamefont
  {Sachdev}},\ }\href@noop {} {\emph {\bibinfo {title} {Quantum Phase
  Transitions}}},\ \bibinfo {edition} {2nd}\ ed.\ (\bibinfo  {publisher}
  {Cambridge University Press},\ \bibinfo {year} {2011})\BibitemShut {NoStop}%
\bibitem [{\citenamefont {Beliaev}()}]{Beliaev1958}%
  \BibitemOpen
  \bibfield  {author} {\bibinfo {author} {\bibfnamefont {S.}~\bibnamefont
  {Beliaev}},\ }\href@noop {} {\bibinfo  {journal} {Zh. {\'E}ksp. Teor. Fiz.
  34, 433 (1958) [Sov. Phys. JETP 7, 299 (1958)]}\ }\BibitemShut {NoStop}%
\bibitem [{\citenamefont {Giorgini}(2000)}]{Giorgini2000}%
  \BibitemOpen
\bibfield  {journal} {  }\bibfield  {author} {\bibinfo {author} {\bibfnamefont
  {S.}~\bibnamefont {Giorgini}},\ }\href {\doibase 10.1103/PhysRevA.61.063615}
  {\bibfield  {journal} {\bibinfo  {journal} {Phys. Rev. A}\ }\textbf {\bibinfo
  {volume} {61}},\ \bibinfo {pages} {063615} (\bibinfo {year}
  {2000})}\BibitemShut {NoStop}%
\end{thebibliography}

%

\end{document}


\title{Supplementary information:\\  Rotation of quantum impurities in the presence of a many-body environment}

\author{Richard Schmidt} 
\email{richard.schmidt@cfa.harvard.edu}
\affiliation{ITAMP, Harvard-Smithsonian Center for Astrophysics, 60 Garden Street, Cambridge, MA 02138, USA}%
\affiliation{Physics Department, Harvard University, 17 Oxford Street, Cambridge, MA 02138, USA} %

\author{Mikhail Lemeshko} 
\email{mikhail.lemeshko@ist.ac.at}
\affiliation{IST Austria (Institute of Science and Technology Austria), Am Campus 1, 3400 Klosterneuburg, Austria}

\begin{abstract}

\end{abstract}

\maketitle

Here we provide some technical details on the derivation of the angulon Hamiltonian, Eq.~(2) of the main text. As described there, we start with the microscopic Hamiltonian describing  the bosons, the impurity, and the impurity-boson interactions: $H =  H_\text{bos} + H_\text{imp} + H_\text{imp-bos}$. The last term, given by Eq.~(1) of the main text, explicitly depends on the orientation of the impurity in the laboratory frame, $(\hat{\theta}, \hat{\phi})$. As a first step, we expand the Hamiltonian in fluctuations around a homogeneous BEC of condensate density $n$ and perform the Bogoliubov transformation, neglecting the term $\sim \bed_\mathbf{k}  \be_\mathbf{k}$, which is suppressed by $1/\sqrt{n}$. In Cartesian coordinates, up to a constant mean-field shift $\sim V(\vec 0) n$, the resulting Hamiltonian reads:
\begin{equation}
\label{HFrohlich}
	H =B \mathbf{J^2} + \sum_\mathbf{k} \omega_k \bed_\mathbf{k}  \be_\mathbf{k} + \sqrt{n} \sum_\mathbf{k} V_\text{imp-bos}(\mathbf{k}, \hat{\theta}, \hat{\phi}) \sqrt{\frac{\epsilon_\mathbf{k}}{\omega_\mathbf{k} }} (\be_\mathbf{k} + \bed_\mathbf{-k}),
\end{equation}
where $\sum_\mathbf{k}  \equiv \int d^3k/(2\pi)^3$.

\subsection*{Impurity-boson interaction}
We now proceed with deriving the expression for $ V_\text{imp-bos}(\mathbf{k}, \hat{\theta}, \hat{\phi}) $. As shown in Fig.~1(a) of the main text, in addition to the laboratory coordinate frame, $(x,y,z)$,  we define the molecular coordinate frame, $(x', y', z')$, needed to describe the microscopic molecule-boson interactions. In the molecular frame, an axially-symmetric interaction potential can be expanded over  spherical harmonics as:
\begin{equation}
\label{Vr}
	V_\text{imp-bos} (\mathbf{r'}) = \sum_\lambda u_\lambda f_\lambda (r') Y_{\lambda 0} (\Theta', \Phi'),
\end{equation}
where $u_\lambda$ and $f_\lambda(r')$ represent, respectively, the strength and shape of the potential in the corresponding angular momentum channel $\lambda$, cf.~Fig.~1(b) in the main text, and  $(r',\Theta', \Phi')$ represent the boson coordinates in the molecular frame. Next we transform Eq.~\eqref{Vr} to the laboratory frame, $(r,\Theta,\Phi)$, using Wigner rotation matrices $D^\lambda_{\mu\nu}$~\cite{VarshalovichAngMom, ZareAngMom}:
\begin{equation}
\label{WignerD}
	Y_{\lambda 0} (\Theta', \Phi') = \sum_\mu D^{\lambda \ast}_{\mu 0} (\hat{\phi}, \hat{\theta}, \hat{\gamma}) Y_{\lambda \mu} (\Theta, \Phi) 
\end{equation}
 Here the Euler angles $(\hat{\phi}, \hat{\theta}, \hat{\gamma})$ define the relative orientation of the two coordinate systems. For a linear molecule, the third angle, $\hat{\gamma}$, can be dropped. Then, using that $D^{\lambda \ast}_{\mu 0} (\hat{\phi}, \hat{\theta}, 0) = \sqrt{\frac{ 4 \pi}{2 \lambda +1}} Y_{\lambda \mu} (\hat{\theta}, \hat{\phi})$~\cite{VarshalovichAngMom, ZareAngMom}, we obtain:
\begin{equation}
\label{Vr2}
	V_\text{imp-bos}(\mathbf{r}, \hat \theta,\hat \phi) = \sum_{\lambda \mu} \sqrt{\frac{ 4 \pi}{2 \lambda +1}}  u_\lambda f_\lambda (r)  Y_{\lambda \mu} (\Theta, \Phi) Y_{\lambda \mu} (\hat \theta,\hat \phi)
\end{equation}
Here $\mathbf{r} \equiv (r, \Theta, \Phi)$ gives the bosons' positions in the laboratory frame.

 The interaction in the momentum space which appears in Eq.~\eqref{HFrohlich} is given by the Fourier transform of Eq.~(\ref{Vr2}), evaluated using the expansion of $e^{-i\veck \vecr}$ in terms of spherical harmonics:
\begin{equation}
\label{Vk}
	V_\text{imp-bos}(\mathbf{k}, \hat \theta,\hat \phi)  \equiv \int d^3\mathbf{r}~V_\text{imp-bos}(\mathbf{r}, \hat \theta,\hat \phi)~e^{-i \mathbf{kr}}= \sum_{\lambda \mu} (2\pi)^{3/2}  i^{-\lambda} \tilde{V}_\lambda (k)  Y_{\lambda \mu} (\Theta_k, \Phi_k) Y_{\lambda \mu} (\hat{\theta}, \hat{\phi}),
\end{equation}
where $\mathbf{k}  \equiv (k, \Theta_k, \Phi_k)$ is the momentum vector in the laboratory frame. This yields the expression $\tilde{V}_\lambda (k) =  u_\lambda 2^{3/2}/\sqrt{2\lambda + 1} \int_0^\infty  dr\, r^2 f_\lambda(r) j_\lambda (kr)$, with $j_\lambda (kr)$ the spherical Bessel function.

\subsection*{Hamiltonian in the angular momentum representation}
\label{AppA}

It is convenient to express the creation and annihilation operators of Eq.~(\ref{HFrohlich}) in the spherical basis using the following relations:
\begin{equation}
\label{AkAklm}
	\bd_\mathbf{k} = \frac{(2\pi)^{3/2}}{k} \sum_{\lambda \mu}  \bd_{k\lambda \mu}~i^{-\lambda}~Y_{\lambda \mu} (\Theta_k, \Phi_k), 
	\end{equation}
\begin{equation}
\label{AklmAk}
	\bd_{k\lambda \mu} =\frac{k}{(2\pi)^{3/2}} \int   d\Phi_k d\Theta_k~\sin\Theta_k~\bd_\mathbf{k} ~i^\lambda~ Y^*_{\lambda \mu} (\Theta_k, \Phi_k)
\end{equation}
Here the quantum numbers $\lambda$ and $\mu$ define, respectively, the angular momentum of the bosonic excitation and its projection onto the laboratory-frame $z$-axis. The transformations defined by Eqs.~(\ref{AkAklm}) and ~(\ref{AklmAk}) correspond to the following commutation relations:
\begin{equation}
\label{AkComm}
	[\be_\mathbf{k}, \bd_\mathbf{k'}] = (2\pi)^3\delta^{(3)}(\mathbf{k-k'})
\end{equation}
\begin{equation}
\label{AklmComm}
	[\be_{k\lambda \mu}, \bd_{k'\lambda' \mu'}] = \delta(k-k') \delta_{\lambda \lambda'} \delta_{\mu \mu'}
\end{equation}

After substituting  Eq.~(\ref{Vk}) and~(\ref{AkAklm}) into Eq.~(\ref{HFrohlich}) and integrating over the angles, the effective Hamiltonian in the spherical basis takes the following form:

\begin{equation}
\label{untransH}
H= B \mathbf{J^2} + \sum_{k \lambda \mu}  \omega_k \bed_{k\lambda \mu} \be_{k\lambda \mu} +   \sum_{k \lambda \mu}  U_\lambda(k)  \left[ Y^\ast_{\lambda \mu} (\hat \theta,\hat \phi) \bed_{k \lambda \mu}+ Y_{\lambda \mu} (\hat \theta,\hat \phi) \be_{k \lambda \mu} \right],
\end{equation}
where $\sum_k \equiv \int dk$ and $U_\lambda(k) = \tilde{V}_\lambda(k) k \sqrt{n \epsilon_\mathbf{k}/\omega_\mathbf{k}} = u_\lambda \left[\frac{8 n k^2\epsilon_k}{\omega_k(2\lambda+1)}\right]^{1/2} \int dr r^2 f_\lambda(r) j_\lambda (kr)$.
\begin{figure}[h]
\centering
 \includegraphics[width=0.01\linewidth]{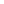}
  \end{figure}


%